\documentclass[final,3p,times]{elsarticle}
\usepackage{epsfig}
\usepackage{color}

\usepackage{amssymb}

\usepackage{lineno}

\begin{document}
\setlength{\unitlength}{1cm}

\begin{frontmatter}



\title{Measurement of Atmospheric Neutrino Oscillations with the ANTARES Neutrino Telescope}





\author[UPV]{S.~Adri\'an-Mart\'inez}
\author[CPPM]{I. Al Samarai}
\author[Colmar]{A. Albert}
\author[UPC]{M.~Andr\'e}
\author[Genova]{M. Anghinolfi}
\author[Erlangen]{G. Anton}
\author[IRFU/SEDI]{S. Anvar}
\author[UPV]{M. Ardid}
\author[NIKHEF]{T.~Astraatmadja\fnref{tag:1}}
\author[CPPM]{J-J. Aubert}
% ...

\fntext[tag:1]{\scriptsize{Also at University of Leiden, the Netherlands}}
\fntext[tag:2]{\scriptsize{On leave of absence at the Humboldt-Universit\"at zu Berlin}}
\fntext[tag:3]{\scriptsize{Also at Accademia Navale de Livorno, Livorno, Italy}}
\fntext[tag:4]{\scriptsize{Deceased}}

\address[UPV]{\scriptsize{Institut d'Investigaci\'o per a la Gesti\'o Integrada de les Zones Costaneres (IGIC) - Universitat Polit\`ecnica de Val\`encia. C/  Paranimf 1 , 46730 Gandia, Spain.}}
\address[CPPM]{\scriptsize{CPPM, Aix-Marseille Universit\'e, CNRS/IN2P3, Marseille, France}}
\address[Colmar]{\scriptsize{GRPHE - Institut universitaire de technologie de Colmar, 34 rue du Grillenbreit BP 50568 - 68008 Colmar, France }}
\address[UPC]{\scriptsize{Technical University of Catalonia, Laboratory of Applied Bioacoustics, Rambla Exposici\'o,08800 Vilanova i la Geltr\'u,Barcelona, Spain}}
\address[Genova]{\scriptsize{INFN - Sezione di Genova, Via Dodecaneso 33, 16146 Genova, Italy}}
\address[Erlangen]{\scriptsize{Friedrich-Alexander-Universit\"at Erlangen-N\"urnberg, Erlangen Centre for Astroparticle Physics, Erwin-Rommel-Str. 1, 91058 Erlangen, Germany}}
\address[IRFU/SEDI]{\scriptsize{Direction des Sciences de la Mati\`ere - Institut de recherche sur les lois fondamentales de l'Univers - Service d'Electronique des D\'etecteurs et d'Informatique, CEA Saclay, 91191 Gif-sur-Yvette Cedex, France}}
\address[NIKHEF]{\scriptsize{Nikhef, Science Park,  Amsterdam, The Netherlands}}
% ...
\address[IRFU/SPP]{\scriptsize{Direction des Sciences de la Mati\`ere - Institut de recherche sur les lois fondamentales de l'Univers - Service de Physique des Particules, CEA Saclay, 91191 Gif-sur-Yvette Cedex, France}}
% ...
\address[Genova-UNI]{\scriptsize{Dipartimento di Fisica dell'Universit\`a, Via Dodecaneso 33, 16146 Genova, Italy}}


\begin{abstract}

The data taken with the ANTARES neutrino telescope from 2007 to 2010, a total live time of 863 days, 
are used to measure the oscillation parameters of atmospheric neutrinos. Muon tracks are reconstructed 
with energies as low as 20 GeV. 
Neutrino oscillations will cause a suppression of vertical upgoing muon neutrinos of such energies crossing the Earth.
The parameters determining the oscillation of atmospheric neutrinos are extracted by fitting 
the event rate as a function of the ratio of the estimated neutrino energy and reconstructed 
flight path through the Earth. Measurement contours of the oscillation parameters in a two-flavour approximation 
are derived. Assuming maximal mixing, a mass difference of 
$\Delta m_{32}^2=(3.1\pm 0.9)\cdot 10^{-3}$ eV$^2$
is obtained, in good agreement with the world average value.

\end{abstract}

\begin{keyword}
neutrino oscillations \sep neutrino telescope \sep ANTARES


\end{keyword}

\end{frontmatter}


\section{Introduction}

A measurement of the quantum mechanical phenomenon of neutrino oscillations provides 
important information on the mass differences of the neutrino mass eigenstates and their mixing angles. 
The effect \textcolor{black}{was discovered by} Super-Kamiokande~\cite{skfirst} 
on neutrinos produced in the Earth atmosphere.
In this paper, the first observation of neutrino oscillations by a high 
energy neutrino telescope is presented. 
As this measurement addresses a higher neutrino energy range, 
\textcolor{black}{the analysis presented here is complementary
to previous measurements. In particular, 
the selection procedure for $\nu_\mu$ events, on which the oscillation parameters are
measured, is different with respect to a similar analysis at lower energies. 
Whereas the separation of $\nu_\mu$ charged current events from $\nu_e$ and neutral
current events becomes much easier, handling the background from misreconstructed
downgoing muons is more challenging (see Section~\ref{selection}).}

The main goal of the ANTARES deep sea neutrino telescope~\cite{antares} is the observation of high energy neutrinos 
from non-terrestrial sources. The telescope is optimised for the detection of Cherenkov light 
induced by the passage of upgoing muons produced in charged current interactions of neutrinos at TeV energies, 
for which the muon can traverse completely the equipped detector volume. 
At lower neutrino energies, both the cross section and the muon range are smaller resulting 
in a decrease of the detection efficiency. 
For the analysis presented here, the lowest energies of detectable muons from neutrino interactions are
about 20 GeV. For upgoing atmospheric neutrinos which 
traverse the Earth, this threshold is low enough for the observed flux of $\nu_\mu$ induced events to be 
significantly suppressed by neutrino oscillations. By studying the observed upgoing muon 
rate as a function of the ratio of the reconstructed muon energy and zenith angle, constraints 
on the atmospheric neutrino oscillation parameters are derived.

The paper is organised as follows: 
in Section 2 the phenomenology of atmospheric neutrino oscillations is introduced and the method used 
to extract the oscillation parameters is discussed. 
The ANTARES telescope is described in Section 3. 
The data set and the Monte Carlo simulations 
are explained in Section 4. 
The methods to reconstruct the neutrino direction and energy are discussed 
in Section 5 followed by details of the event selection in Section 6. 
A discussion on systematic 
uncertainties is given in Section 7 and the final results are presented in Section 8.

\section{Neutrino Oscillations}
\label{sec:fit}
The survival probability of atmospheric $\nu_\mu$ in the framework of three flavour mixing is given as
\begin{equation}
  P(\nu_\mu \rightarrow \nu_\mu) = 
  1 - 4|U_{\mu 1}|^2|U_{\mu 2}|^2 \sin^2 \biggl({1.27 \Delta m_{21}^2 L \over E_\nu} \biggr)
  - 4|U_{\mu 1}|^2|U_{\mu 3}|^2 \sin^2 \biggl({1.27 \Delta m_{31}^2 L \over E_\nu} \biggr)
  - 4|U_{\mu 2}|^2|U_{\mu 3}|^2 \sin^2 \biggl({1.27 \Delta m_{32}^2 L \over E_\nu} \biggr)
\label{eq:3flavour}
\end{equation}
where $L$ is the travel path (in km) of the neutrino through the Earth and $E_\nu$, its energy (in GeV).
$U_{\alpha i}$ is the $3\times 3$ \textcolor{black}{PMNS-}matrix which describes the mixing
between flavour eigenstates $\nu_e,\nu_\mu,\nu_\tau$ and mass eigenstates 
$\nu_1,\nu_2,\nu_3$ and $\Delta m_{ij}^2 = |m_i^2 - m_j^2|$ (in eV$^2$) 
is the absolute difference of the squares of the masses of the corresponding neutrino mass eigenstates.
\textcolor{black}{The present analysis is restricted to $L<12800~$km and $E_\nu>20$~GeV, for which 
the term $\sin^2 (1.27 \Delta m_{21}^2 L/E_\nu)$ does not exceed $3.7\cdot 10^{-3}$ when using 
$\Delta m_{21}^2$ from~\cite{pdg}. This term can safely be ignored as well as differences between 
$\Delta m_{31}^2$ and $\Delta m_{32}^2$ and  Equation~\ref{eq:3flavour} simplifies to}
\begin{equation}
 P(\nu_\mu \rightarrow \nu_\mu) = 
 1 - 4\left(1-|U_{\mu 3}|^2\right)|U_{\mu 3}|^2 \sin^2 \biggl({1.27 \Delta m_{32}^2 L \over E_\nu} \biggr)
\label{equation:dm2-domin}
 \end{equation} 
\textcolor{black}{Results could in principle be extracted in terms of $|U_{\mu 3}|^2$ and 
$\Delta m_{32}^2$ 
which are the two oscillation parameters in Equation~\ref{equation:dm2-domin}. 
To maintain compatibility with earlier results, a mixing angle $\sin^2\theta_{23} = |U_{\mu 3}|^2$
is defined, ignoring the 2.4\% deviation from 1 of $\cos^2\theta_{13} = 0.976$~\cite{daya-bay}. 
This leads to the usual two-flavour description}
\begin{equation}
 P(\nu_\mu \rightarrow \nu_\mu) = 
  1 - \sin^2 2\theta_{23} \sin^2 \biggl({1.27 \Delta m_{32}^2 L \over E_\nu} \biggr)
 =1 - \sin^2 2\theta_{23} \sin^2 \biggl({16200~ \Delta m_{32}^2 \cos\Theta \over E_\nu} \biggr). 
\label{equation:vacuum}
\end{equation}
For upgoing tracks $L$ is in good approximation related to the zenith angle $\Theta$ by 
$L=D\cdot\cos\Theta$ where $D$ is the Earth diameter. 
The transition probability, $P$, depends now on only
two oscillation parameters, $\Delta m_{32}^2$ and $\sin^2 2\theta_{23}$,
which determine the behaviour for the atmospheric neutrino oscillations.

With $\Delta m_{32}^2=2.43\cdot 10^{-3}$~eV$^2$ and $\sin^2 2\theta_{23}=1$ from~\cite{pdg}
one expects the first oscillation maximum, {\it i.e.} $P(\nu_\mu \rightarrow \nu_\mu)=0$
for vertical upgoing neutrinos ($\cos\Theta=1$) of $E_\nu=$24~GeV. 
Muons induced by 24 GeV neutrinos can travel up to 120 m in sea water.

The observed number of events in bin $i$, $N_i$, of a given variable 
can be compared to the number $MC_i$ of expected Monte Carlo events in the same channel
\begin{equation}
MC_i = \mu_i + \sum_k 
\left[1 - \sin^22\theta_{23}~\sin^2\left(\frac{16200~ \Delta m^2_{32} \cos\Theta_{ik}}{E_{\nu,{ik}}}\right)\right]
\label{eq:mc}
\end{equation} 
where $\mu_i$ is the number of the background atmospheric muon events in channel $i$ 
and the sum gives the number
of atmospheric neutrino events in channel $i$ weighted by the event dependent oscillation 
probability from Equation~\ref{equation:vacuum}.
As the oscillation probability $P(\nu_\mu \rightarrow \nu_\mu)$ 
depends on $E_\nu/\cos\Theta$, the natural choice for a variable in which the channel $i$
can be defined
is the ratio between a quantity which depends on the neutrino energy 
and the reconstructed zenith angle, $\Theta_R$. As explained in Section~\ref{reco}, 
the energy-dependent variable is the observed muon range in the detector.
The oscillation parameters are extracted by a $\chi^2$ minimisation which is detailed in 
Section~\ref{sec:syst}.

\section{The ANTARES Detector}

A detailed description of the ANTARES detector 
can be found in~\cite{antares}.
The detector consists of 12 lines, equipped with photosensors,
and a junction box
which distributes the power and
clock synchronization signals to the lines and collects the data.
The junction box is connected to the shore by a 42~km
electro-optical cable.
The length of the detection lines is 450~m, of which the lowest 100~m are not instrumented.
Their horizontal separation
is about 65~m and they are arranged to form a
regular octagon on the sea floor.
They are connected
to the junction box with the help of a submarine using wet-mateable
connectors.
Each line comprises 25 storeys each separated by a vertical distance of 
14.5~m. The lines are kept taut by
a buoy at the top of the line and an anchor on the seabed.
The movement of the line elements due to the sea currents is continuously measured
by an acoustic calibration system with an accuracy of 10~cm~\cite{acoust}.

Each storey contains three 45$^\circ$
downward-looking 10'' photomultiplier tubes (PMT) inside
pressure resistant glass spheres - the optical modules~\cite{om}.
Some of the storeys contain supplementary calibration equipment such as acoustic hydrophones
or optical beacons~\cite{led-beacon}.

The signals of each photomultiplier are readout by two
ASICs. The charges and arrival times of the PMT signals
are digitised and stored for
transfer to the shore station~\cite{ars}. 
The time stamps are synchronised by a clock signal which is sent
at regular intervals from the shore to all electronic cards.
The overall time calibration is better than 0.5~ns~\cite{timing}. 
Therefore the
time resolution of the signal pulses is limited by the
transit time spread of the photomultipliers ($\sigma \sim $1.3~ns)~\cite{pmt} and by chromatic dispersion
for distant light sources.
All data are sent to the shore station.
With the observed optical background rate of 70~kHz per PMT at the single photon level this
produces a data flow of several Gbit/s to the shore.
In the shore station a PC farm performs a data filtering to reduce
the data rate by at least a factor of 100~\cite{daq}. Several trigger 
algorithms are applied depending
on the requested physics channel and on the observed optical noise.

\section{Data and Simulations}


The present analysis is based on data taken with the ANTARES detector
between March 2007 and December 2010. Until December 2007 ANTARES operated 
in a 5-line configuration, followed by several months of
operation with 10 installed detector lines. The detector construction was 
completed in May 2008. 
All physics runs taken under normal conditions have been used. 
The events selected by two tight trigger conditions are used~\cite{antares}.
The analysed sample consists of 293 million triggers, 
dominated by atmospheric muons, corresponding to a detector live time of 863 days. 


Downgoing atmospheric muons were simulated with the program MUPAGE~\cite{mupage,bundle}
which provides parametrised muon bundles at the detector. 
Upgoing neutrinos were simulated according to the parametrisation of 
the atmospheric $\nu_\mu$ flux from~\cite{bartol} in the energy range from 10~GeV to 10~PeV. 
The Cherenkov light, produced inside or in the vicinity of the detector instrumented volume, 
was propagated taking into account
light absorption and scattering in sea water~\cite{light}.
The angular acceptance, quantum efficiency and other characteristics of the PMTs
were taken from~\cite{om} and the overall geometry corresponded to the layout of the
ANTARES detector~\cite{antares}. The optical noise was simulated from counting rates observed 
in the data. At the same time, the definition of
active and inactive channels has been applied from data runs as well.
The generated statistics corresponds to an equivalent observation time of 100~years 
for atmospheric neutrinos and ten months for atmospheric muons.

\section{Reconstruction}
\label{reco}

The algorithm used for the muon track reconstruction is described in~\cite{bbfit}.
It assumes a simplified detector geometry composed of straight vertical detection lines.
The method combines a strict selection of direct Cherenkov photon hits which are grouped around 
``hot spots" at each detector line with a $\chi^2$-like fitting procedure. A hot spot
corresponds to a signal of at least 4~photoelectrons seen on two adjacent storeys of the same detector
line within a narrow time window of less than 100~ns. Only hits on detector lines with 
such a hot spot are used in the track fitting. If the selected hits occur only on one
detector line, a single-line fit is performed. No azimuth angle is determined in this case due
to the rotational symmetry of the problem. This does not affect the present measurement, 
as the oscillation probability does not depend on 
the azimuth angle (see Equation~\ref{equation:vacuum}).
If selected hits occur instead on several detector lines, a multi-line fit is performed which
provides both the zenith and azimuth angles of the track.
The inclusion of single-line events is a special feature of the reconstruction 
method and allows to significantly lower the energy threshold of the final
atmospheric neutrino sample. Whereas for multi-line events the threshold energy of the
final neutrino sample is about 50~GeV due to the 65~m horizontal gap between lines, 
single-line events are reconstructed 
down to 20~GeV for nearly vertical tracks, accessing events with $E_\nu/L$ values close to
the first oscillation maximum. 

\begin{figure}[htbp]
    \begin{picture}(8,6)
      \epsfclipon 
      \epsfxsize=8cm
      \put(4,0){\epsfbox{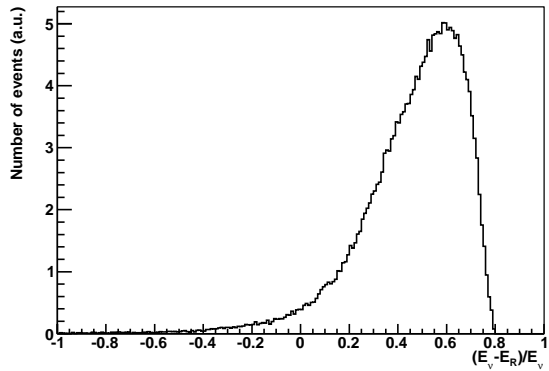}}
    \end{picture}
\caption{Difference between the true neutrino energy, $E_\nu$, and $E_R$ normalised by $E_\nu$ 
for low energy events ($E_\nu<100$~GeV) from the final
event sample of Section~\ref{selection}.}
\label{fig:Enu_Er}
\end{figure}

The neutrino energy is estimated from the observed muon range in the detector. 
The selected hits are sorted according to their vertical position, $z$.
The z-coordinates of the uppermost and lowermost hits, $z_{max}$ and $z_{min}$, together with 
the reconstructed zenith angle, $\Theta_R$, allow to define an approximate muon range
\begin{equation}
 S = (z_{max}-z_{min})/\cos\Theta_R.
\label{muonrange}
\end{equation}
When considering the geometry of the Cherenkov light cone, Equation~\ref{muonrange} 
is exact only for strictly vertical tracks. It has been verified that the use of a more 
sophisticated range definition does not improve the precision of the measurement for 
the oscillation parameters. $S$ is used to estimate the muon energy by taking 
into account the ionisation energy loss of 0.2~GeV/m for minimum ionising muons in sea water~\cite{pdg}
\begin{equation}
E_R = S \cdot 0.2~\textrm{GeV/m}.
\label{Er}
\end{equation}
The derived energy from the visible muon range in the detector can be considered
as a lower limit of the actual neutrino energy. 
The presence of a hadronic shower at the neutrino vertex is ignored, as well as  
the fact that the muon might leave or enter the detector, making only a fraction 
of its actual range available for measurement.
The normalised difference between the true neutrino energy, $E_\nu$, and $E_R$ is shown in
Figure~\ref{fig:Enu_Er} for all selected simulated events (see Section~\ref{selection})
with $E_\nu<100$~GeV. The mean value of 0.45 illustrates the fact that $E_R$ measures on average 
about half of the neutrino energy. The RMS of the distribution is 0.22.

\section{Event Selection}
\label{selection}

Downgoing atmospheric muons \textcolor{black}{dominate the ANTARES event sample on the trigger level
as illustrated in Table~\ref{tab:reduction}. Neutrinos contribute less than $10^{-4}$ here.
In principle a simple cut in the reconstructed zenith angle should be enough to separate both
classes. But as seen from Table~\ref{tab:reduction}, this is not sufficient. Atmospheric muons are often
seen as bundles and they can be accompanied by hard stochastic processes such as bremsstrahlung. 
Both effects complicate their correct reconstruction and a certain fraction of downgoing atmospheric muon 
events are misreconstructed as upgoing.}
Some cuts on the quality of the reconstructed tracks are needed to reduce this contamination and derive reliably 
neutrino oscillation parameters from the data set.
The goodness of the track fit is measured by the ``normalised fit quality" as introduced in~\cite{bbfit},
a quantity equivalent to a $\chi^2$ per number of degrees of freedom (NDF).
The selection cuts, which are described below, have been obtained
from a blind analysis. The single-line data sample has been kept blind, thereby masking
a possible oscillation signature.

For the multi-line selection, only events which have hits on more than 5 storeys are kept to
allow a non-degenerate track fit. Further, the fit must not converge on a physical boundary 
of any of the fit parameters. 
As the contamination of misreconstructed atmospheric muons is particularly strong close 
to the horizon, a further condition, $\cos\Theta_R>0.15$, is imposed, {\it i.e.} 
tracks closer than $9^\circ$ to the horizon are excluded. 

\begin{figure}[htbp]
    \begin{picture}(8,6)
      \epsfclipon 
      \epsfxsize=8cm
      \put(0.5,0){\epsfbox{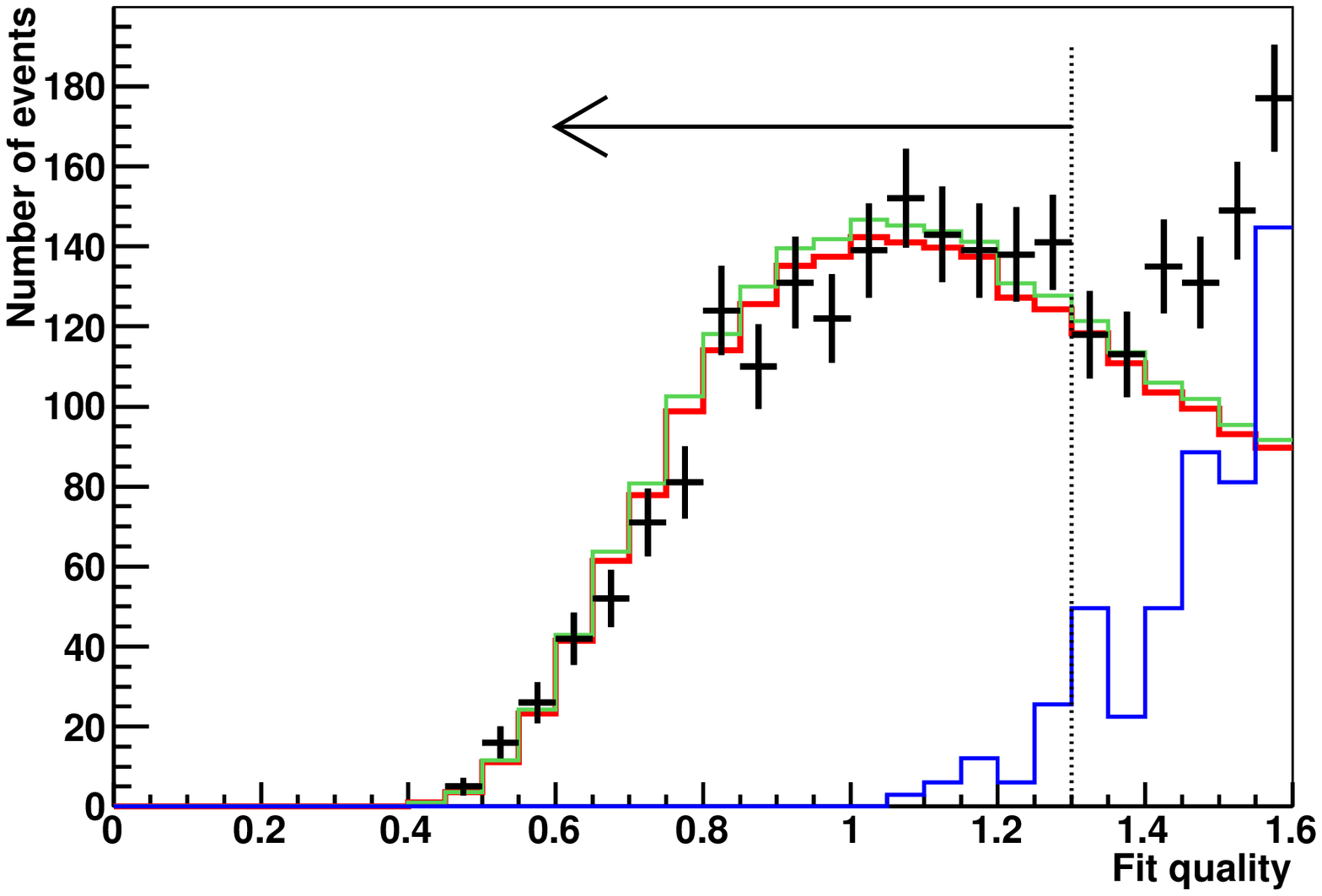}}
      \put(8.5,0){\epsfbox{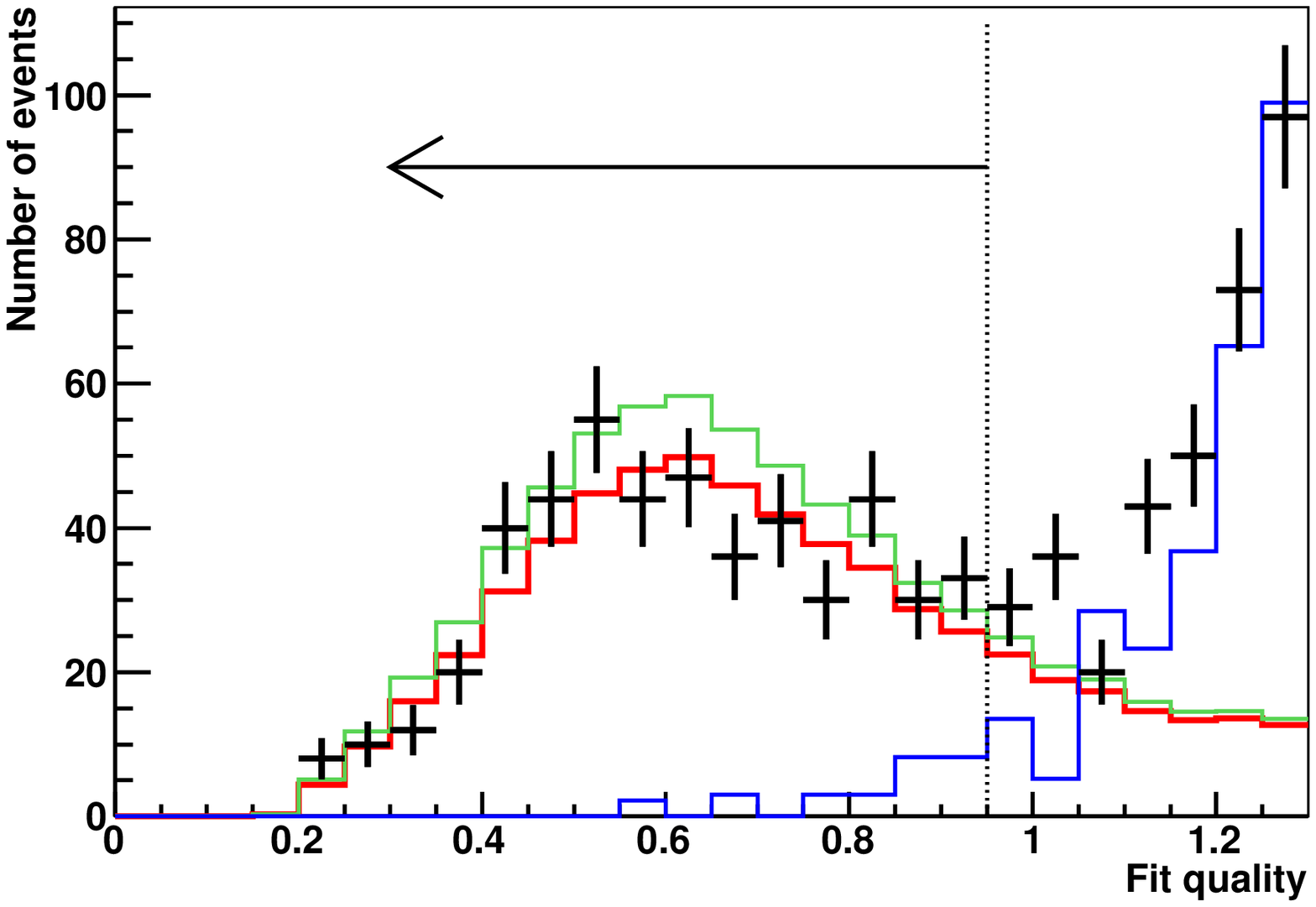}}
    \end{picture}
\caption{Normalised fit quality of the final multi-line (left) and single-line (right) samples.
Data with statistical errors (black) are compared to simulations from atmospheric neutrinos 
with oscillations assuming parameters from~\cite{pdg} (red) and without oscillations (green) and atmospheric muons (blue). 
For a fit quality larger than 1.6 (multi-line) or 1.3 (single-line) 
the misreconstructed atmospheric muons dominate. 
The arrows indicate the chosen regions.}
\label{fig:tchi2}
\end{figure}
The distribution of the normalised track fit quality of the resulting multi-line event sample 
for data and simulations is shown in Figure~\ref{fig:tchi2} (left). 
The neutrino Monte Carlo 
samples are scaled down by an overall normalisation factor $r=0.86$ as obtained from the fit
(see Section~\ref{sec:results}).
\textcolor{black}{This is well within the uncertainty of the atmospheric neutrino
flux model~\cite{fluxerror}.}
Figure~\ref{fig:tchi2} (left) shows that a 
cut on the normalised fit quality allows to cleanly separate the upgoing neutrinos from the downgoing 
muons. 
In order to have a contamination of misreconstructed atmospheric muons below 5\%, 
a cut value of $1.3$ is chosen. 

For the single-line selection, events which have hits on more than 7~storeys are kept. This
yields a minimal track length for a vertical upgoing muon of about 100~m, which can be
produced by a muon of 20~GeV. Further cuts are identical to the multi-line selection.

The distribution of the normalised track fit quality of the resulting single-line event sample 
for data and simulations is shown in Figure~\ref{fig:tchi2} (right). The neutrino Monte Carlo 
samples are again scaled down by a factor $r=0.86$. Figure~\ref{fig:tchi2} (right) 
shows that also for this data set a cut on the normalised fit quality allows to 
cleanly separate the downgoing muons from the 
upgoing neutrinos. 
In order to have a contamination of misreconstructed atmospheric muons below 5\%, 
a cut value of $0.95$ is chosen. 
 
\begin{table}[htpb]
\begin{center}
\begin{tabular}{||c|c|c|c|c|c|c||} \hline
        & \multicolumn{3}{c|}{Multi-line} 
	& \multicolumn{3}{c||}{Single-line}\\ \hline
                      & Data             & $\nu$ MC & $\mu$ MC      &  Data            & $\nu$ MC & $\mu$ MC \\ \hline 
All                   & $1.42\cdot 10^8$ &  8755 & $1.23\cdot 10^8$ & $1.51\cdot 10^8$ &  8242 & $1.10\cdot 10^8$ \\
Nstorey $>$ N$_{cut}$ & $1.33\cdot 10^8$ &  8248 & $1.18\cdot 10^8$ & $4.44\cdot 10^7$ &  1260 & $3.03\cdot 10^7$ \\
Fit boundary          & $1.32\cdot 10^8$ &  8150 & $1.17\cdot 10^8$ & $4.31\cdot 10^7$ &  1242 & $2.93\cdot 10^7$\\
$\cos\Theta_R>0.15$   & $2.74\cdot 10^6$ &  5512 & $1.84\cdot 10^6$ & $7.97\cdot 10^5$ &  1116 & $6.96\cdot 10^5$\\ \hline
Fit quality cut & $ 1632\pm 40$    & $1971\pm 6$ & $52\pm 12$ &  $494\pm 22$     & $651\pm 3$ & $28\pm 9$\\ 
                             &     & $1910\pm 6$ &                           &   & $557\pm 3$ & \\ \hline
\end{tabular}
\end{center}
\caption{Event reduction due to the cuts used. Statistical errors are given for the final data set. 
The effect of oscillations with parameters from~\cite{pdg} is taken into account 
only for the values given in the very last row.}
\label{tab:reduction}
\end{table}
The effect of the different selection cuts in the two channels is detailed in Table~\ref{tab:reduction} 
for data and the Monte Carlo sets. Satisfactory agreement between data and Monte Carlo numbers is
observed at all cut levels.

Events from $\nu_e$ charged current (CC) interactions as well as neutral
current (NC) interactions produce cascade-like event topologies.
\textcolor{black}{The spatial extension of these cascades, which are composed of the 
hadronic and electromagnetic
showers at the neutrino interaction vertex, does not exceed more than a few meters. 
This is in contrast 
to the lowest energetic selected muons, which have still a range of at least 100~m 
in sea water and therefore a 
distinct topology. The above described selection cuts, 
tuned to suppress the atmospheric muon background, reduce efficiently also the contribution from
$\nu_\mu$ NC and $\nu_e$ interactions.}
Their contribution to the final event sample is estimated to be less than one event 
in the multi-line channel and 6 events in the single-line channel.
\textcolor{black}{Both numbers are significantly smaller than the residual background contribution from
atmospheric muons}.

The zenith angle of the final neutrino sample is reconstructed with a precision of
$0.8^\circ$ for multi-line events and $3.0^\circ$ for single-line events 
(median of the angular error distribution with respect to the true neutrino direction). 
\begin{figure}[htbp]
    \begin{picture}(8,6)
      \epsfclipon 
      \epsfxsize=8cm
      \put(4,0){\epsfbox{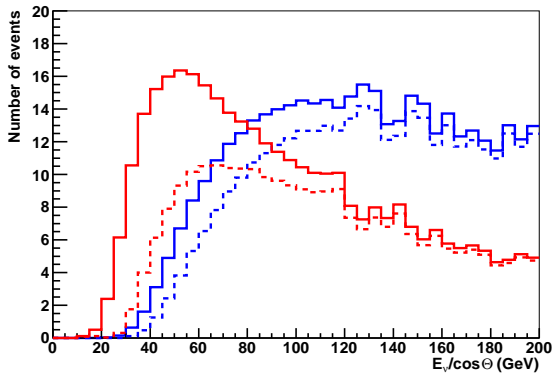}}
    \end{picture}
\caption{Distribution of $E_\nu/\cos\Theta$ for the selected events of the atmospheric neutrino simulation.
The solid lines are without neutrino oscillations, 
the dashed lines include oscillations assuming the best fit values reported in~\cite{pdg}. 
The red histograms indicate the contribution of the single-line sample, in
blue the multi-line events.}
\label{fig:EL_MC}
\end{figure}
The distribution in $E_\nu/\cos\Theta$ for the simulated selected atmospheric neutrino samples 
is shown in Figure~\ref{fig:EL_MC}. 
The expected event numbers are shown for the actual detector live time of 863 days.
The effect of neutrino oscillations is seen by comparing the Monte Carlo curves with and 
without oscillations. The importance of the inclusion of the single-line events 
is evident. 
The expected deficit due to neutrino
oscillations for $E_\nu/\cos\Theta<200$~GeV amounts to 143~events, 
the majority of which (91~events) are in the single-line sample. These oscillated events
are mainly converted into $\nu_\tau$ and might produce muons by a $\nu_\tau$ CC interaction
followed by a decay $\tau^-\rightarrow\mu^-\bar\nu_\mu\nu_\tau$ (or charge conjugate). 
To estimate the appearance of muons 
from $\nu_\tau$ interactions in the final event sample, the energy dependent cross section ratio 
$\sigma(\nu_\tau CC)/\sigma(\nu_\mu CC)$, about 0.5 for $E_\nu=25$~GeV, the 17\% branching ratio of the 
muonic $\tau$-decay and the soft energy spectrum of the resulting muons have to be considered.
It is found that the expected deficit due to oscillations is reduced by less than 5~events.


\section{Systematic Uncertainties}
\label{sec:syst}

Systematic uncertainties lead to a correlated distortion of the $E_R/\cos\Theta_R$ distribution. 
They are implemented as pull factors in the
$\chi^2$ function to be minimised:
\begin{equation}
\chi^2 = \sum_i\left[N_i - (1+\epsilon) MC^{1L}_i-(1+\eta) MC^{ML}_i\right]^2/\sigma_i^2
+ (\epsilon-\eta)^2/\sigma_R^2.
\label{chi2}
\end{equation}
The sum extends over the bins in $E_R/\cos\Theta_R$ with $N_i$ 
the observed number of events in bin $i$ whereas $MC^{1L}_i$ and $MC^{ML}_i$ denote the expected 
event numbers from simulations in the single-line and multi-line channel respectively and 
$\sigma_i$ is the corresponding statistical error of bin $i$.
The simulated event numbers depend on the oscillation parameters according to Equation~\ref{eq:mc}.
The quantities $\epsilon$ and $\eta$ are two pull factors which allow to renormalise the two channels.
The variations of $\epsilon$ and $\eta$ are constrained by $\sigma_R$, which is introduced below.

A large class of uncertainties modify the overall normalisation, {\it i.e.} they act on $\epsilon$ and $\eta$
in the same way.
The normalisation of the neutrino flux model as well as the uncertainty
on the absolute neutrino detection efficiency of the ANTARES detector
fall into this class. Equation~\ref{chi2} does not constrain such uncertainties, as indicated by the absence of terms
such as $\epsilon^2/\sigma_\epsilon^2$ or $\eta^2/\sigma_\eta^2$,
{\it i.e.} the relative overall normalisation between data and simulations is left totally free in this
analysis.

A second class of uncertainties might change the shape of the $E_R/\cos\Theta_R$ distribution.
Single-line events have typically lower $E_\nu/\cos\Theta$ values
(and thus correspondingly lower observed $E_R/\cos\Theta_R$ values) 
than multi-line events (see Figure~\ref{fig:EL_MC}). 
A change in the ratio of the total number of events in each sample, $R_0=N^{1L}/N^{ML}$, 
would therefore reflect a shape change of the 
$E_R/\cos\Theta_R$ distribution. In terms of the pull factors this can be written as
$R(\epsilon,\eta)=R_0(1+\epsilon)/(1+\eta)\approx R_0(1+\epsilon-\eta)$. Therefore $R(\epsilon,\eta)$
varies with $\epsilon-\eta$ if both pull factors are small.

Several simulations have been performed with modified input parameters which may
affect in a slightly different manner vertical and horizontal events (thereby acting on $\cos\Theta_R$) 
as well as low and high energy events~\cite{antsyst}. 
The average quantum efficiency of the phototubes was changed by $\pm$10\% 
as well as their angular acceptance.
The water absorption length of sea water was also varied by $\pm$10\%.
Further, the cuts in the normalised fit quality were varied in two steps of 0.05 around 
the chosen values thereby testing the stability of the analysis procedure. 
The spectral index of the atmospheric neutrino flux was varied by $\pm 0.03$ as suggested
in~\cite{ICsyst}.
As a result a set of values of $R$ was derived.
It was observed that $R$ remains stable within 5\% of its original value, $R_0$. 
This value $\sigma_R=0.05$ is used in Equation~\ref{chi2} to constrain relative variations of $\epsilon$ and $\eta$.
This accounts for the maximal variation of the shape of the $E_R/\cos\Theta_R$ 
distribution due to the considered systematic effects.


\section{Results}
\label{sec:results}
\begin{figure}[htbp]
    \begin{picture}(8,6)
      \epsfclipon 
      \epsfxsize=8cm
      \put(0.5,0){\epsfbox{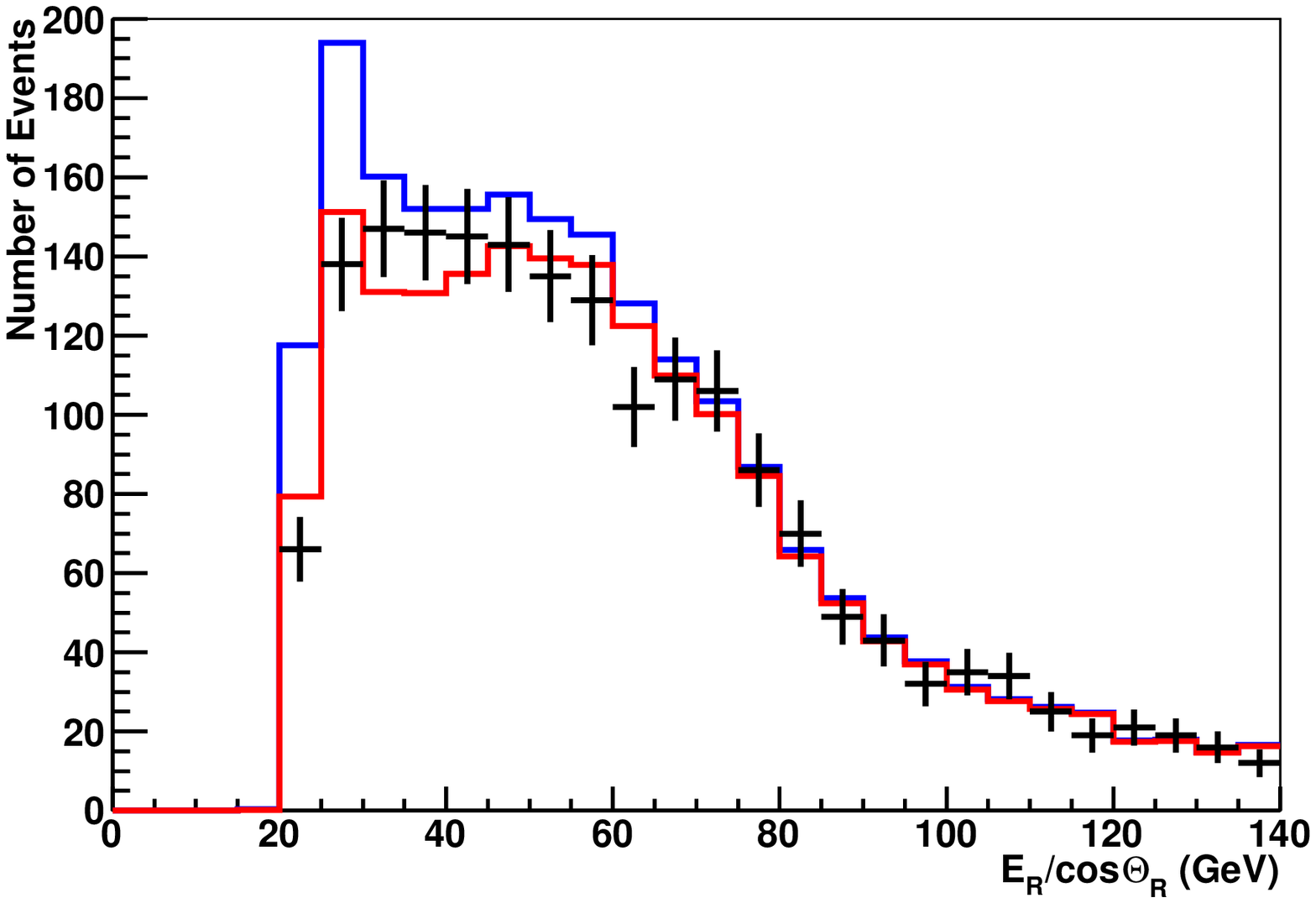}}
      \put(8.5,0){\epsfbox{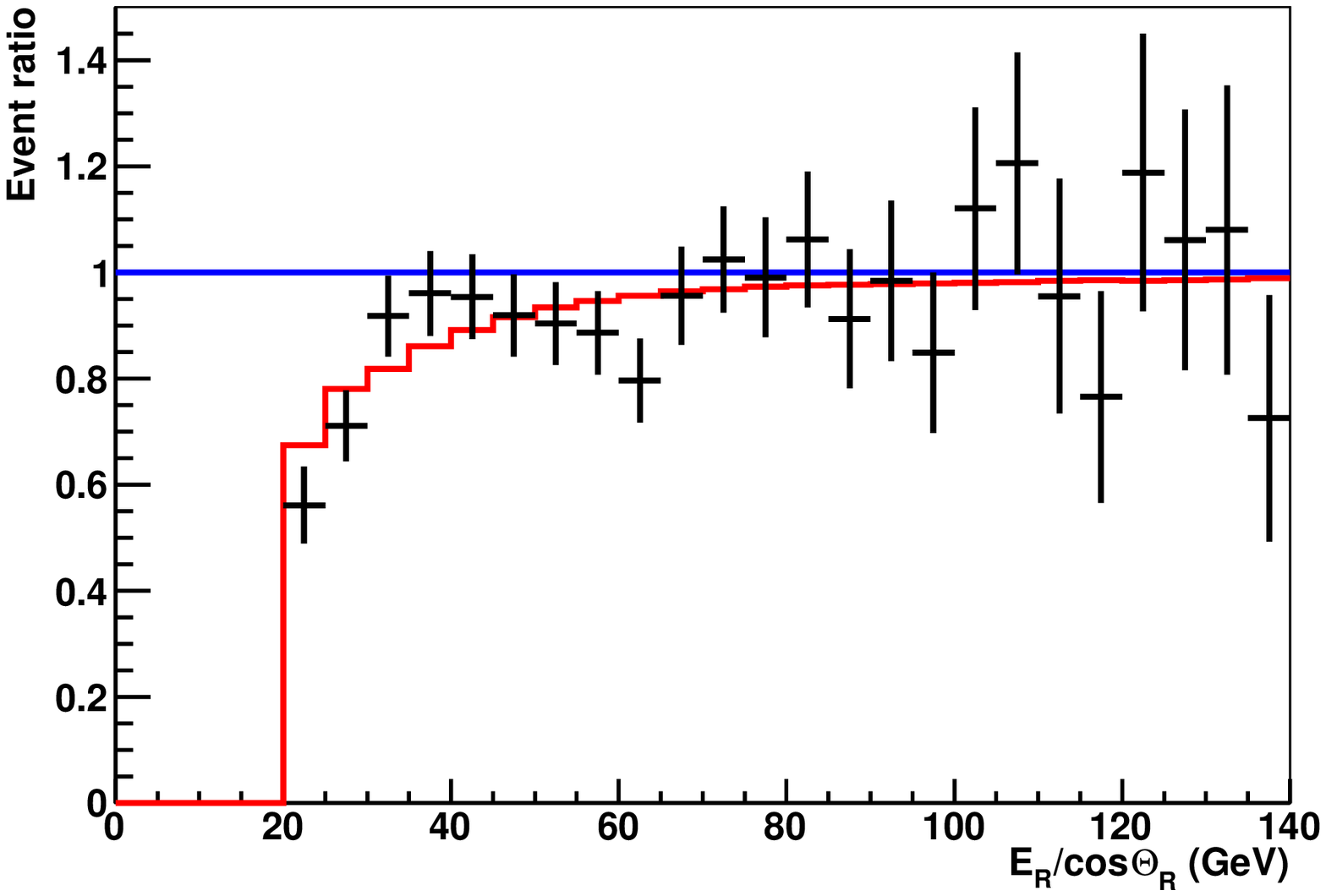}}
    \end{picture}
\caption{{\bf Left}: Distribution of $E_R/\cos\Theta_R$ for selected events. Black crosses
are data with statistical uncertainties, whereas the blue histogram 
shows simulations of atmospheric neutrinos without neutrino oscillations (scaled down by a factor 0.86)
plus the residual background from atmospheric muons. 
The red histogram shows the result of the fit.
{\bf Right}: The fraction of events
with respect to the non-oscillation hypothesis. Same color code as for the left figure.}
\label{fig:EL_fit}
\end{figure}

After applying the cuts defined in Section~\ref{selection} 
the final event numbers obtained are given in Table~\ref{tab:reduction}.
Figure~\ref{fig:EL_fit} (left) shows the resulting $E_R/\cos\Theta_R$ distribution 
for data and simulations.
Figure~\ref{fig:EL_fit} (right) shows the fraction of measured and simulated events
with respect to the non-oscillation Monte Carlo hypothesis and indicates a 
clear event deficit for $E_R/\cos\Theta_R<60$~GeV, 
as expected assuming atmospheric neutrino oscillations. 
Restricting the $\chi^2$ minimisation to parameters in the physically allowed region
yields the red curve of Figure~\ref{fig:EL_fit} with $\Delta m^2_{32} = 3.1\cdot 10^{-3}\mbox{eV}^2$
and $\sin^22\theta_{23} = 1.00$.
The corresponding pull factors are $\epsilon=-0.138$ and $\eta=-0.142$.
This is used as an overall normalisation parameter $r=0.86$ in Figures~\ref{fig:tchi2}~and~\ref{fig:EL_fit}.
The fit is performed in 25 bins: the 24 bins shown in 
Figure~\ref{fig:EL_fit} plus one overflow bin which contains $299$ events with 
$E_R/\cos\Theta_R>140$~GeV. The fit yields $\chi^2/\mbox{NDF} = 17.1/21$. 
When imposing the world average oscillation parameters,
$\Delta m_{32}^2=2.43\cdot 10^{-3}$~eV$^2$ and $\sin^2 2\theta_{23}=1$ from~\cite{pdg}, ~$\chi^2/\mbox{NDF} = 18.4/21$
is found.

For the non-oscillation hypothesis, 
{\it i.e.} $\sin^22\theta_{23}=0$, $\chi^2/\mbox{NDF} =31.1/23$ is obtained.
The pull factors in this case are $\epsilon=-0.302$ and $\eta=-0.196$. 
The event deficit in the single-line channel is seen here as 
$\epsilon$ becoming lower than $\eta$. Requiring in addition $\epsilon=\eta$ the $\chi^2$ increases further to 
$\chi^2/\mbox{NDF}=40.0/24$, which has a probability of only 2.1\%.



\begin{figure}[htbp]
    \begin{picture}(8,8)
      \epsfclipon 
      \epsfxsize=12cm
      \put(2,0){\epsfbox{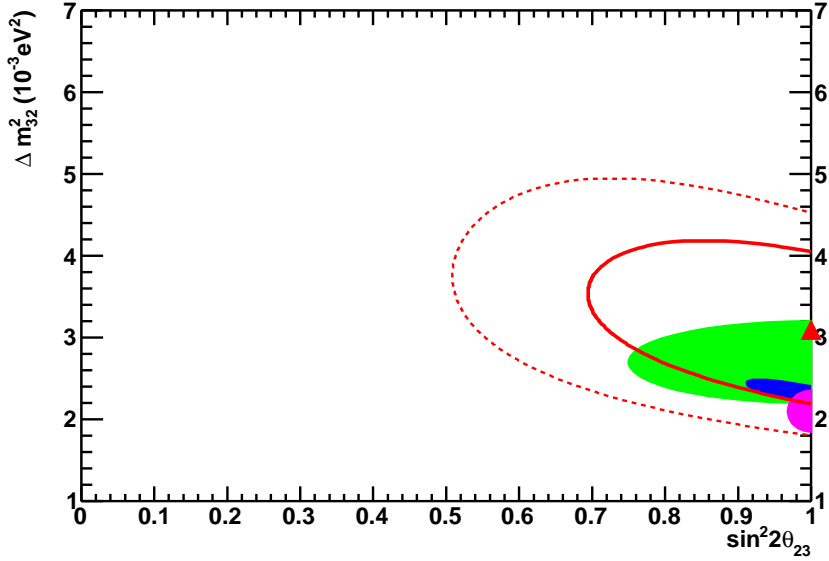}}
    \end{picture}
\caption{\textcolor{black}{68\% and 90\% C.L. contours (solid and dashed red lines)} 
of the neutrino oscillation parameters as derived from the 
fit of the $E_R/\cos\Theta_R$ distribution. The best fit point is indicated by the triangle.
The solid filled regions show results at 68\% C.L. from K2K~\cite{k2k} (green), MINOS~\cite{minos} 
(blue) and Super-Kamiokande~\cite{superk} (magenta) for comparison.}
\label{fig:oscil}
\end{figure}

This measurement is converted into contours of the
oscillation parameters and is shown in Figure~\ref{fig:oscil}. 
\textcolor{black}{With 68\% C.L. $\sin^2 2\theta_{23}>0.70$ is found and
$\Delta m^2_{32}$ is constrained to values in the range $[2.2, 4.2]\cdot 10^{-3} \mbox{eV}^2$.}
If maximal mixing is imposed ($\sin^2 2\theta_{23}=1$), 
the obtained range of $\Delta m^2_{32}$ is
\begin{equation}
\Delta m^2_{32}=(3.1\pm 0.9)\cdot 10^{-3} \mbox{eV}^2.
\end{equation}
The results are in agreement with other measurements from K2K~\cite{k2k}, MINOS~\cite{minos} 
and Super-Kamiokande~\cite{superk}.

\section{Conclusions}


Based on data taken by the ANTARES neutrino telescope from 2007 to 2010, constraints on 
the neutrino oscillation parameters $\sin^2 2\theta_{23}$ and $\Delta m^2_{32}$ have been derived. 
If maximal mixing is assumed, a value 
$\Delta m^2_{32}=(3.1\pm 0.9)\cdot 10^{-3} \mbox{eV}^2$ is obtained. 
The result agrees well with current world data and demonstrates a good understanding of 
the performance of the ANTARES telescope at its lowest accessible energies. 
It is the first such measurement by a high energy neutrino telescope 
\textcolor{black}{and underlines the potential of future low energy extensions 
of the existing neutrino telescopes for such physics}.

\section{Acknowledgements}

The authors acknowledge the financial support of the funding agencies:
Centre National de la Recherche Scientifique (CNRS), Commissariat
\`{a} l'\'{e}nergie atomique et aux \'{e}nergies alternatives  (CEA), 
Commission Europ\'{e}enne (FEDER fund and Marie Curie Program), 
R\'{e}gion Alsace (contrat CPER), R\'{e}gion
Provence-Alpes-C\^{o}te d'Azur, D\'{e}\-par\-tement du Var and Ville de
La Seyne-sur-Mer, France; Bundesministerium f\"{u}r Bildung und Forschung
(BMBF), Germany; Istituto Nazionale di Fisica Nucleare (INFN), Italy;
Stichting voor Fundamenteel Onderzoek der Materie (FOM), Nederlandse
organisatie voor Wetenschappelijk Onderzoek (NWO), the Netherlands;
Council of the President of the Russian Federation for young scientists
and leading scientific schools supporting grants, Russia; National
Authority for Scientific Research (ANCS - UEFISCDI), Romania; Ministerio
de Ciencia e Innovaci\'{o}n (MICINN), Prometeo of Generalitat Valenciana
and MultiDark, Spain; Agence de l'Oriental and CNRST, Morocco. We also
acknowledge the technical support of Ifremer, AIM and Foselev Marine
for the sea operation and the CC-IN2P3 for the computing facilities. 





\bibliographystyle{elsarticle-num}



\end{document}